# Benefits and Barriers to Human-AI Teaming for Crisis Counseling

Ivy Pham*[1], Akshay Swaminathan*[2], Sebastian Garcia*[3], Jaimie Lim[4], Gloria Ye[5,6], Divyanjali Verma[7], Alan Pham[8], Chastin Chung[5], Shreya Shah[9], Shaked Azzam, Sharang Phadke[10], Ivan Lopez[2], William Wang[11,12], Angelina Wang[13], Nigam H. Shah[5,6]

[1] University of California, Davis, School of Medicine, Sacramento, CA, USA
[2] Department of Biomedical Data Science, Stanford School of Medicine, Stanford, CA, USA
[3] University of Wisconsin School of Medicine and Public Health, Madison, WI, USA
[4] Department of Biostatistics and Computational Biology, University of Rochester, Rochester, NY, USA
[5] Department of Medicine, Stanford School of Medicine, Stanford, CA, USA
[6] Clinical Excellence Research Center, Stanford School of Medicine, Stanford, CA, USA
[7] Vandrevala Foundation, India
[8] Stanford University, Stanford, CA, USA
[9] The Ohio State University, Columbus, OH, USA
[10] Center for Biomedical Informatics Research, Stanford University, Stanford, CA, USA
[11] Keck School of Medicine, University of Southern California, Los Angeles, CA, USA
[12] California Institute of Technology, Pasadena, CA, USA
[13] Cornell Tech, New York, NY, USA

*Equal contribution

Corresponding Author: Akshay Swaminathan (akshaysw@stanford.edu)

# Abstract

### Background

AI systems are increasingly deployed to support mental health clinicians, yet real-world evidence remains limited. In crisis counseling, where clients often seek human connection, understanding whether AI augmentation preserves or undermines meaningful human engagement is essential.

### Methods

In a mixed-methods study of a counselor-facing AI copilot on a WhatsApp-based crisis hotline in India (Mar.-Oct. 2025), counselors could request, edit, and send AI messages. We analyzed 20 interviews, 8 shadowing sessions, 56 survey responses, and edits in 27,778 copilot messages from 108 counselors via thematic analysis (Technology Acceptance Model 2, plus inductive coding of emergent factors), and survey statistics.

### Results

Counselors reported high copilot acceptance (survey): 98.2% found it helpful; 89.3% likely to continue use; 80.4% reported efficiency gains. Interviews highlighted sentence scaffolding, empathetic phrasing, stress reduction, catching missed details, and efficiency. Barriers (survey) included repetitive outputs (23.2%), lack of context (17.9%), and authenticity concerns (10.7%). While 69.6% (survey) reported skill enhancement, interviews surfaced skill atrophy concerns. Counselors frequently accepted copilot suggestions, editing to remove questions, adjust tone, or personalize style.

### Conclusion

In one of the largest real-world deployments of a counselor-facing mental health AI copilot, counselors reported reduced cognitive load, empathetic communication, and greater efficiency, alongside concerns about skill atrophy, authenticity, and over-reliance. In settings like crisis counseling where clients seek human connection, AI tools must balance operational support with preserving clinical skill and genuine therapeutic engagement. Designs soliciting approval or editing of complete drafts may overweight efficiency at the expense of sustained human engagement.

# Introduction

Global demand for mental health providers far outstrips supply, especially in crisis settings where timely care is critical. Crisis counselors face high stress and cognitive load, juggling concurrent text conversations, managing suicide risk, and maintaining therapeutic focus, conditions that drive burnout and turnover.[1–6] AI-powered chatbots are increasingly proposed as scalable solutions with the potential to extend access, personalize interventions, and reduce clinician load.[7–9] Despite promising early outcomes, reviews caution that evidence remains limited, particularly regarding long-term clinical effectiveness, safety, and user engagement.[10,11] Naive adoption of fully automated systems risks undermining therapeutic authenticity, mishandling crisis nuance, or producing unsafe responses. In contrast, AI augmentation to assist rather than replace counselors has emerged as a more viable path, with recent work showing that AI can scaffold and improve counselor skills.[12,13]

Understanding counselors' perspectives is essential for evaluating AI adoption, as their judgments determine whether new tools become trusted aids. Clients are also turning to AI chatbots for mental health support, with early evidence of both benefits and risks.[14–16] Here, we focus on the complementary perspective of providers, whose views have been studied in relation to perceptions of hypothetical or experimental systems. Surveys and qualitative studies highlight both optimism about efficiency and skepticism regarding therapeutic authenticity, safety, and professional identity.[17–20] Very few studies have examined counselors' lived experiences with AI systems deployed in practice, a consequential gap in high-stakes crisis counseling.

To fill this gap, we conducted a mixed-methods investigation into the deployment of a counselor-facing AI "copilot" within a WhatsApp-based crisis hotline in India. The system was deployed organization-wide to more than 200 counselors and embedded into live, synchronous crisis conversations, supporting over one million counselor messages across more than 200,000 conversations while preserving counselor control over all outgoing messages. The copilot can be invoked at any point during active conversations to generate draft responses that counselors could edit, send, or discard. Drawing on 20 in-depth counselor interviews, 8 shadowing sessions, 56 survey responses, and edit-pattern analysis of 27,778 counselor-edited messages from 108 counselors using the copilot, we analyze the real-world factors shaping how counselors accept or reject AI, and how AI reshapes counseling practices. This study offers one of the first empirical looks at how crisis counselors encounter and navigate AI assistance in live real-world practice. In doing so, it highlights critical considerations for designing augmentation tools that support counselors while preserving the integrity of therapeutic relationships.

# Methods

## Study Setting and Participants

We conducted this study at a WhatsApp-based crisis counseling hotline in India providing free, 24/7 mental health support. Beginning on March 8, 2025, an AI copilot was integrated into the hotline's chat interface as a human-in-the-loop assistant. Counselors could choose when to activate it to generate a draft message aligned with organizational training guidelines and crisis counseling best practices. Drafts were displayed in a separate panel and could be accepted as written, edited, or rejected altogether, ensuring that clinical judgment and responsibility remained with the human counselor. The system was explicitly framed as a tool to support, not replace, counselors, with safeguards to minimize harmful outputs. Participants included volunteer and full-time counselors with varying experience (<6 months to >3 years) and educational backgrounds (82.14% of survey respondents with psychology post-graduate degrees). This study was determined to not qualify as human subjects research by WCG IRB.

## Data Collection

We used a mixed-methods approach: (1) 20 semi-structured interviews (30-60 minutes) exploring usage patterns, editing behaviors, workflow adaptations, and professional identity impacts; (2) 8 real-time shadowing sessions (~30 minutes) observing copilot usage during live crisis conversations; (3) 56 survey responses (response rate = 81%, out of 69 counselors who used the copilot ≥1 month) using 5-point Likert scales and open-ended questions to assess factors like usefulness, efficiency, stress reduction, skill development, and ethical concerns; and (4) 27,778 counselor-edited copilot messages drawn from a corpus of 94,209 messages sent by 108 counselors using the copilot within 7 months after deployment. Interview and shadowing participants were selected non-randomly to cover a wide range of copilot usage/non-usage and counselor demographics.

## Coding Approach

We conducted thematic analysis[21] on data from interviews, shadowing sessions, and open-ended survey questions using a hybrid deductive-inductive approach. Our codebook contained 5 themes and 21 primary codes with subcategories. Three coders from a team of six independently coded each transcript at the paragraph level, allowing multiple codes per segment, and iteratively refining the codebook through team discussions. A code was counted for a participant if at least one of the three coders applied it.

Independent coding agreement was modest ($\kappa = 0.438$ per-code; $\kappa = 0.482$ per-document). After implementing a two-rater resolution process for disputed segments, negotiated agreement increased (post-consensus $\kappa = 0.914$ per-code; $\kappa = 0.919$ per-document) (Supplementary Tables 2-3). No codes remained conflicting ($\kappa < 0.4$).

We integrated quantitative and qualitative analyses by using metrics to establish prevalence of themes, and thematic analysis and quotes to contextualize patterns.

## Analysis of Edits

We analyzed paired data consisting of LLM-generated suggestions and corresponding counselor-edited responses. We started with all 94,209 messages sent by counselors using the copilot from March 8, 2025 (date of copilot deployment) until October 8, 2025. We excluded messages where counselors rejected the copilot’s suggestion, as well as messages where no substantial edits were made (70.5%), defined as a token-level BLEU score above 0.9 between the suggestion and the final response. We selected this threshold based on manual inspection, which confirmed that pairs with BLEU > 0.9 typically reflected only very minor edits (e.g., small punctuation or wording tweaks) rather than substantive rewriting. This yielded 27,778 messages used for the edit analysis.

We used an LLM to categorize the types of edits counselors made. A prompt instructed the model to compare each AI-generated suggestion with the counselor's edited response and assign one or more category labels: question removal, question modification, increased empathy, decreased empathy, increased specificity, decreased specificity, or other (Appendix G). The model produced structured JSON output containing the assigned label(s) and a brief justification for each classification. The LLM we used was OpenAI Chat Completions API (gpt-4.1) in a batched pipeline, with generation length limited to 500 tokens per message and temperature parameter set to 0.1. Model outputs were validated through manual review of a random subset. We report descriptive statistics summarizing the frequency of each edit type along with representative examples and mean BLEU scores.

# Results

## Drivers of Acceptance

We analyzed counselor acceptance through the lens of the Technology Acceptance Model 2 (TAM2),[22] which posits technology adoption via perceived usefulness and ease of use, supplemented by social and cognitive factors. Among 56 survey respondents, 98.2% rated the copilot as helpful or extremely helpful, with 89.3% likely or extremely likely to continue using it and 80.4% agreeing or strongly agreeing that it is easy to use (Figures 1-2, Supplementary Table 4).

### Communication Support: Phrasing, Empathy, and Language

Language scaffolding was a primary driver of acceptance. Survey data showed 83.9% agreed or strongly agreed the copilot helps them "structure and frame sentences better" (Figure 2, Supplementary Table 4). Interviews similarly highlighted sentence scaffolding and paraphrasing as key benefits (Supplementary Table 7).

The copilot's ability to generate empathetic responses was also the benefit most often volunteered in open-ended survey answers (28.6%), and it was repeatedly highlighted in interviews. One counselor explained: "I tend to sometimes speed up… and forget the empathy part... this reminds me to bring in that empathy" (Interview: Counselor I5).

The copilot also handles communication barriers effectively, such as understanding a client with broken English (Shadowing: Counselor S1) or providing English answers even if clients use Hindi (Interview: Counselor I9).

**Cognitive Support During High-Pressure Situations**

The intense emotional demands of crisis counseling emerged vividly. One counselor described the work as "extremely emotionally exhausting…especially with... suicide cases... [or] past trauma" (Interview: Counselor I11), even needing to pause while discussing this topic during the interview, illustrating how deeply crisis work affects counselors. Similarly, others echoed exhaustion from handling multiple simultaneous chats and lengthy trauma narratives.

In response, 78.6% of survey respondents agreed or strongly agreed that they experienced reduced stress when using the copilot (Figure 2, Supplementary Table 4), a benefit also reflected in interviews. It may provide relief through cognitive scaffolding. As one counselor described: "[The copilot] gives a moment to think... it shifts the emotional load from a counselor to AI" (Interview: Counselor I11). Cognitive overload was a common cue for use among interviewees (85.0%), and 14.3% of survey respondents mentioned it in open-ended answers.

**Efficiency and Caseload Management**

Among survey respondents, 80.4% agreed or strongly agreed that the copilot improves overall efficiency (Figure 2, Supplementary Table 4). Benefits included increased case capacity, better management of simultaneous cases, and faster response speed.

**Back-up Support**

The copilot functions as a safety net, enhancing counselor effectiveness by catching important details and offering fresh perspectives in real-time. Interviewees described instances where the copilot surfaced details they had missed. One described that it caught a client's husband's absence, thus their “whole perspective changed” (Interview: Counselor I3). Another said it helped them to explore a mother-client relationship that they "didn't pick up," and they praised the copilot as "a nice protective shield" (Interview: Counselor I5). Additionally, 19.6% of survey respondents reported turning to the copilot when feeling stuck.

**Learning Support**

Overall, 69.6% of survey respondents agreed or strongly agreed that the copilot enhanced their counseling skills, and interviews noted that it could reinforce training protocols. The degree to which counselors agreed that the copilot enhances counseling skills decreased with experience, from a mean of 3.90 (<6 months experience) to 3.77 (6-12 months) to 3.75 (1-3 years) to 3.57 (>3 years) on a 5-point scale.

Some frequent AI users emphasized thoughtful engagement. One counselor stressed developing "a personal ethical orientation to AI” to ensure “learning [is] happening” (Interview: Counselor I1). This counselor discussed using another AI tool to create quiz questions and practice counseling through role-plays, noting how it has improved their knowledge, productivity, efficiency, and enjoyment of learning, which illustrates broader AI capabilities for professional development.

## Barriers to Acceptance

### Response Quality Issues

The most-mentioned limitations in open-ended survey answers were repetitive responses (23.2%) and context blindness (17.9%); 29 of 56 respondents mentioned at least one limitation, and 17 of 56 answered “none” to the barriers question. The copilot's premature solution focus also created challenges: "I've seen Copilot jumping to [coping mechanisms] way too soon" (Interview: Counselor I3) and "The copilot option usually directs the session towards a solution oriented approach and does not allow enough space for the client to open up" (Survey).

### Learning and Skill Development Concerns

Concerns about skill development and over-reliance surfaced mainly in interviews, with some mentions in the survey (7.1% disagreed with perceived skill enhancement; 3.6% mentioned skill development concerns in open-ended answers). One non-user explained that they wanted to learn crisis counseling skills themselves, believing that AI use prevents real learning (Interview: Counselor I2). Others worried about transition challenges: "If I rely on this too much, then I will not really be able to develop my own skills when I need to shift to calls" (Interview: Counselor I7).

### Client Detection of AI and Authenticity Concerns

Counselors (16.1% in surveys) mentioned instances of clients questioning whether they were AI, and interviews described counselors reducing copilot usage after such occurrences. For context, and measured per conversation rather than per counselor, at this same helpline, clients suspected AI use in 1.5% of conversations that involved no AI assistance, rising from 0.8% in June 2024 to 2.6% in March 2025.[23]

Additionally, counselors (21.4% survey) observed that the copilot sounded robotic/AI-like, and interviews echoed similar concerns. Some counselors hypothesized that AI detection triggers include immediate response times, perfect grammar, and formal phrasing.

They also expressed authenticity worries (10.7% survey), which were raised by most interviewees (90.0%). As one counselor reflected: "They're still not my sentences... I'm not sure if I feel like I've been completely authentic" (Interview: Counselor I15).

In interviews, counselors emphasized the "human touch" as essential in counseling. Some viewed AI as supportive, while others expressed concerns about replacement; some articulated mixed feelings. Some counselors felt "50-50" about the future of AI, feeling anxiety about AI replacing counselors, but also acknowledging the need for human connections (Interview: Counselor I11).

### Ethical Barriers

Some counselors (14.3% survey) had ethical concerns. Confidentiality was raised mainly in interviews (50.0%) and was coded in open-ended answers for 7.1% of survey respondents. Reassurance that client data remains confidential and not used for AI training alleviated this. Still, this highlights how ethics must always be considered in AI tool development. One counselor also cited broader societal ethical concerns: "The environmental damages caused by Gen AI are enough for me to not use it" (Survey).

### Technical Issues

Occasional technical problems were reported (8.9% survey), including generation failures, slow loading, blank responses, and less accurate content when handling multiple cases simultaneously. Interviews also described technical issues. While technical improvements were implemented throughout the deployment period, these issues highlight the need for robust system performance in crisis contexts where response delays could erode counselor trust in the copilot.

## Transformation of Counseling Practice

### Edit patterns

Across 27,778 edited messages, the most common edit was question removal, occurring in 55.0% of cases (for all messages with this edit type, mean BLEU = 0.518), followed by removing empathy statements (33.5%; mean BLEU = 0.475) and modifying questions (25.7%; mean BLEU = 0.528) (Table 1). Counselors also made messages more specific in 16.5% of edits (mean BLEU = 0.500) and added empathy phrases in 7.5% (mean BLEU = 0.479).

From interviews, many counselors reported actively editing copilot responses. Common edits included modifying or removing questions, adjusting tone, and adapting language to match personal style. Shadowing showed some counselors often removed or softened early questions to foster empathy, establish rapport, and allow clients to guide the dialogue (Shadowing: Counselors S3 and S8). Still, other counselors valued the copilot’s questioning, highlighting individual preferences.

### High-Severity Cases

Survey data showed 67.9% agreed or strongly agreed the copilot helps ensure client safety, yet in open-ended answers about where it was most and least helpful (Q3), 19.6% named it most helpful for suicidal or non-suicidal self-injury (NSSI) cases while 14.3% named it least helpful for these same cases.

Perceptions varied by experience: only 45.0% of counselors with <6 months experience endorsed the copilot's safety role, compared to 84.6% (6–12 months), 75.0% (1–3 years), and 85.7% (>3 years). Interviews highlighted this divide: one counselor (>3 years experience; often manages up to 3 clients simultaneously) noted that the copilot could support assessments by suggesting additional safety checks (Interview: Counselor I20); while another counselor (<6 months experience, focuses on one client at a time) remarked, “In high risk cases... I’ll be fully focused on the client... I can write a better response than waiting for the copilot” (Interview: Counselor I3). Survey comments echoed concerns about delays from editing, less accurate ladder-up responses for severe cases, and limited utility in detecting suicidality.

These findings suggest that counselor experience and workflow strongly shape how the copilot is integrated into high-risk situations.

## Discussion

In this large, real-world deployment of a counselor-facing AI copilot in crisis counseling, we found high acceptance alongside consistent, active mediation of AI outputs by counselors. Across surveys, interviews, shadowing, and message-level edit analysis, counselors overwhelmingly described the copilot as valuable for language scaffolding, cognitive offloading, and efficiency during high-pressure moments, while retaining the final say over AI suggestions. When counselors edited drafts, they mostly modified or removed questions and adjusted tone or empathy statements to match their personal style. At the same time, counselors articulated unresolved concerns around authenticity and skill development.

Recent critiques of human-in-the-loop AI designs argue that they constrain scalability[24,25] and undercut the efficiency gains promised by automation.[26] While this may be true in some domains, such critiques often overlook the role of client preferences and trust in mental health care. Prior work at this same helpline found that clients increasingly suspect they may be speaking to an AI, and 21.5% of those who did explicitly stated a preference for a human, suggesting that full automation may undermine engagement even if technically feasible.[23] In such settings, human-in-the-loop approaches may be necessary not as a transitional compromise, but as a design requirement to balance AI’s operational benefits with client expectations.[27,28] Evidence on performance further underscores this nuance: some studies find that human–AI collaboration performs worse than AI alone but better than humans alone,[29] whereas others show that human–AI systems outperform either independently.[30] Although we did not evaluate AI-only counseling in this study, and our findings primarily reflect counselor preferences and self-reported benefits rather than client outcomes, prior work in this same crisis hotline demonstrated that a human-plus-AI copilot improved counseling quality across 23 distinct metrics, supporting the feasibility of augmentation in practice-critical, trust-sensitive environments.

The copilot's impact on professional skill development revealed a paradox. Most counselors reported perceived skill enhancement, yet more than half of interviewees expressed concern about deskilling—a tension well supported by prior literature on automation bias and over-reliance.[31–33] Notably, perceived skill enhancement declined with counselor experience, with novices reporting greater benefits than more experienced counselors. There are limits to using self-reporting as a tool: self-reports can diverge from measured performance,[26,31] and routine AI assistance has been linked to declines in clinicians’ unassisted skill.[34] This raises a broader question facing clinical training: to what extent should clinicians use AI to develop skills versus preserving independent competence? Parallel debates have emerged in medicine around AI-generated documentation, where educators warn against premature reliance before trainees build baseline note-writing skills, while proposing structured guardrails so AI serves as a scaffold rather than a substitute for clinical reasoning.[35,36] In crisis counseling, our findings similarly suggest that the resolution lies not in restricting AI use outright, but in deliberately positioning AI as a tool for skill building rather than substitution. Approaches such as practicing with simulated clients, receiving structured feedback on counseling technique, or using AI to reflect on alternative responses may mitigate deskilling while preventing disengagement. By supporting higher-order clinical reasoning and interpersonal skills while offloading routine cognitive burdens, such models may allow counselors to remain engaged and "practice at the top of their license," fostering collaboration rather than replacement.[33]

Our findings suggest that the current copilot design, where counselors are asked to accept or edit complete draft messages, may prioritize efficiency at the expense of counselor engagement. This raises the question of how to design tools that preserve the cognitive and emotional benefits counselors value (reduced stress, language scaffolding, catching missed details) while sustaining the meaningful human engagement that justifies having a human there at all. Some AI tools in mental health have explored alternative approaches, such as providing targeted suggestions or clinical nudges rather than complete drafts,[37] though it remains unclear whether such designs yield comparable efficiency gains. Potential approaches also include limiting copilot invocations per conversation, restricting copilot use to specific phases of the interaction, or offering component-level suggestions (e.g., empathy statements or safety flags)[38] rather than full draft messages. A triage-based model may offer a path forward, especially considering recent increasing evidence that people are comfortable turning to AI tools for mental health needs.[39,40] Clients who are willing to engage with an autonomous AI counselor could be routed to a fully automated pathway with human escalation available as needed, while clients who prefer or require human connection are directed to a human counselor. This parallels triage models proposed in medicine where AI manages more routine cases and clinicians focus on more complex ones, with the added consideration that in mental health, patient preference must inform routing decisions. In the autonomous pathway, efficiency gains are achieved through automation itself. In the human pathway, the reduced volume pressure allows AI tools to be redesigned around engagement and skill preservation rather than speed, thereby prioritizing nudges, component-level suggestions, or training support over complete draft generation. Whether such a model is feasible, and how to design the AI tools within each pathway, remain open empirical questions and critical priorities for future work.

This study has several limitations. First, findings are drawn from a single organization providing text-based crisis counseling in India, which may limit generalizability to other mental health settings, modalities (e.g., voice or in-person care), or cultural contexts. Second, survey responses were obtained from a subset of counselors (81% response rate), raising the possibility of selection bias. Third, the cross-sectional design captures perceptions and behaviors during deployment but does not assess long-term effects on counselor skill development, burnout, or practice patterns. Fourth, our findings reflect counselor preferences and self-reported benefits, which may not match objectively measured performance;[26] we did not measure client outcomes such as symptom reduction, engagement, or satisfaction. Counselor acceptance is necessary but not sufficient evidence that augmentation improves care. Future work could incorporate anonymized client feedback scores or brief post-session surveys and use validated measures (e.g., a Working Alliance Inventory-style instrument adapted for chat) to operationalize relational constructs such as perceived authenticity and therapeutic alliance. Fifth, survey percentages derived from open-ended answers reflect unprompted mentions and are likely lower bounds. Finally, we did not compare human–AI collaboration with AI-only counseling, limiting conclusions about the relative effectiveness of different delivery models.

In one of the largest real-world deployments of a counselor-facing AI copilot in crisis mental health care, we find near-universal clinician acceptance. Yet acceptance is a necessary but insufficient condition for responsible deployment. Clients in crisis who choose to speak with a human counselor are owed a clinician who is engaged, competent, and present. Our findings expose a fundamental tension: the cognitive and emotional benefits that drive adoption—language scaffolding, stress reduction, efficiency—may gradually erode the very qualities that make human involvement warranted. Future work must determine how AI tools can be designed to enhance efficiency while preserving meaningful human engagement over time. This is not only a technical design problem but an ethical obligation to the individuals who place their trust in human connection in moments of vulnerability.

## Tables and Figures

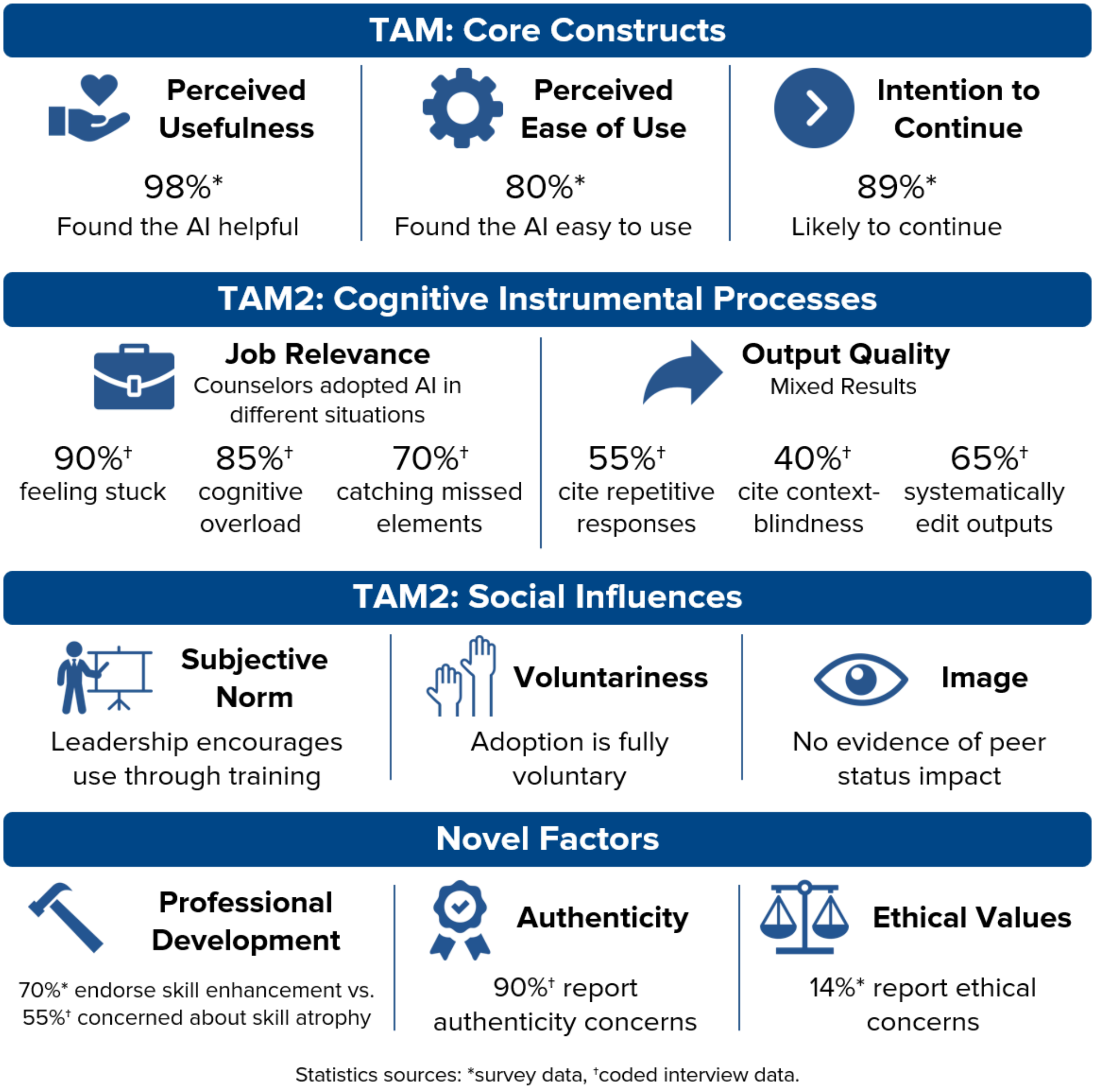

**Figure 1. Summary of key findings organized by Technology Acceptance Model 2 (TAM2) constructs and novel emergent factors.** Core TAM constructs (top) show high perceived usefulness, ease of use, and intention to continue. Cognitive instrumental processes (second row) reflect situational adoption patterns and mixed perceptions of output quality. Social influences (third row) describe the organizational context of adoption. Novel factors (bottom) capture three themes that emerged inductively beyond the TAM2 framework: professional development, authenticity, and ethical values. Statistics are drawn from survey data (*) and coded interview data (†). Sample sizes: survey (n=56), interviews (n=20).

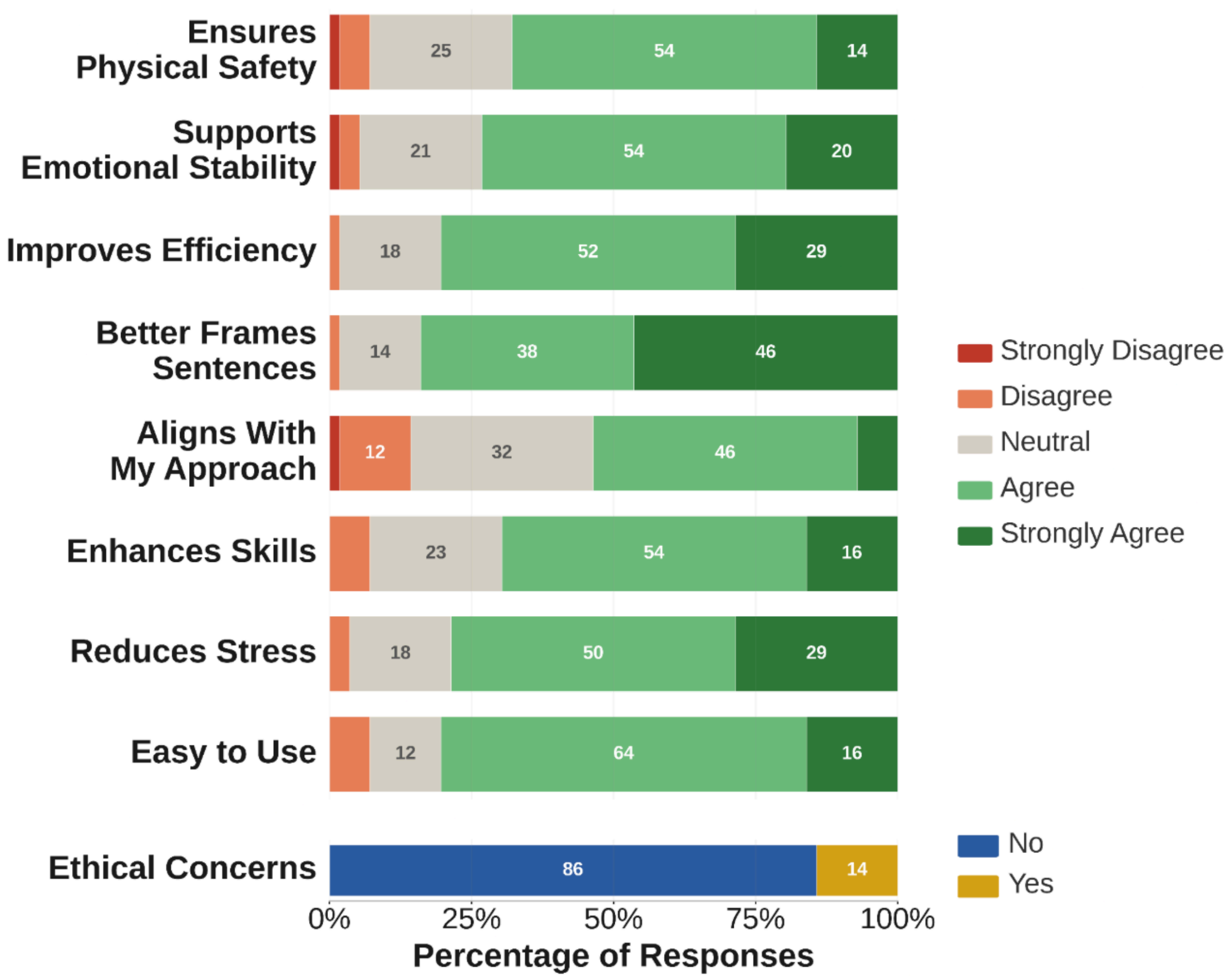

**Figure 2. Counselor survey responses on perceptions of the AI Copilot (n=56).** Stacked horizontal bars show the distribution of Likert-scale responses (Strongly Disagree to Strongly Agree) across nine perception items. The bottom bar displays responses to a binary item on ethical concerns, with 86% of respondents reporting no ethical concerns. Percentages shown within bars reflect the proportion of respondents selecting each response category.

| Edit analysis | | | |
|---|---|---|---|
| **Edit type** | **N (%)** | **Mean BLEU score** | **Example edit** |
| Remove question | 14856 (54.98%) | 0.518 | It sounds like being single at this stage of your life adds to your feelings of uncertainty and loneliness. I can only imagine how this must weigh on your mind with this big change. ~~Is there anything else about your move or being single that's been difficult for you?~~ |
| Edit question | 6950 (25.72%) | 0.528 | I see that this situation is affecting your concentration and emotional stability.~~ Sharing more about your relationship challenges might bring some clarity. Would you like to talk about what's been challenging with your wife?~~ **What would you like to focus on working for this session today?** |
| Add empathy | 2017 (7.46%) | 0.479 | **I get these thoughts are disturbing.** Exploring self-compassion could lead to more self-acceptance and peace. Nurturing kindness toward your feelings can aid healing and growth. What are your thoughts on trying small ways to practice self-compassion each day? |
| Remove empathy | 9038 (33.45%) | 0.475 | ~~I see the strength it took you to share all of that with me. Your experience and feelings are important, and we're here to support you through this.~~ **I know its not fair, its like** you're facing a lot with panic attacks, confusion, and feeling frozen. Have you been able to talk to anyone else about what you're going through, or found anything that helps you even a little? |
| More specific | 4464 (16.52%) | 0.500 | ~~Thank you for sharing that with me.~~ I understand that you are facing challenges with your gender identity. **I read your earlier msgs too,** Can you tell me more about what you've been feeling or any specific challenges you are experiencing right now? |
| Less specific | 15 (0.06%) | 0.490 | I hear that getting lost in daydreaming can affect your focus on important things ~~like your studies~~. Can you tell me more about ~~when you find yourself daydreaming the most~~ **it**? |
| Other | 962 (3.56%) | 0.697 | I'm still here, Aditi, and I'm listening to you. **There seems some network issue.** It's good to know that reading novels provides some relief. It's okay if it doesn't solve everything. Having a small plan for tough moments can be helpful. Is there another activity or mindset that might bring you comfort? |

**Table 1. Classification of counselor edits to AI copilot suggestions.** Edit categories, frequencies, mean BLEU scores, and representative examples across 27,778 edited messages. In the example edits, strikethrough indicates text removed by the counselor and **bold** indicates text added. Categories were assigned using an LLM-based classification pipeline and are not mutually exclusive; a single message could receive multiple labels, so percentages sum to greater than 100%. Lower BLEU scores indicate more extensive modification of the original suggestion. Any client names are fictitious.

## Additional Information

**Funding Statement:**

- AS reported funding from Knight Hennessy Scholars and the PD Soros Fellowship.

**COI Statement:**

- AS reported employment at Benchmark Health and owning equity in Benchmark Health, Daybreak Health, and Cerebral.
- WW reported employment at Coherence Health and owning equity in Coherence Health.
- AP reported employment at Microsoft and owning equity in Microsoft.
- Dr. Shah reported being a cofounder of Prealize Health (a predictive analytics company) as well as Atropos Health (an on-demand evidence generation company) and serving on the Board of the Coalition for Healthcare AI (CHAI), a consensus-building organization providing guidelines for the responsible use of artificial intelligence in health care. Dr. Shah serves as an advisor to Opala, Curai Health, JnJ Innovative Medicines and AbbVie pharmaceuticals.

**Author Contributions:**

- Conceptualization: IP, AS, DV
- Data acquisition: IP, SG, JL, GY, DV, SS, SA, WW
- Data analysis: IP, AS, SG, JL, GY, AP
- Writing and editing: IP, AS, SG, GY, DV, AP, JL, CC, SS, SP, IL, WW, AW
- Supervision: AW, NHS

**Keywords:** AI Copilot, Mental Health, Human-AI Collaboration, Crisis Counseling

# APPENDIX

## Appendix A: Shadowing and Interviewing Demographics

| Counselor ID | Employment Status | Experience Level | Educational Background | # Copilot generations |
|---|---|---|---|---|
| S1 | Full-time Employee | 1-3 years | Psychology Masters Degree | >10,000 |
| S2 | Full-time Employee | 1-3 years | MPhil in Psychology | 100-500 |
| S3 | Full-time Employee | 1-3 years | Psychology Masters Degree | 100-500 |
| S4 | Volunteer | 6-12 months | Psychology Bachelors Student | 501-1000 |
| S5 | Volunteer | 6-12 months | Psychology Bachelors Degree | 501-1000 |
| S6 | Full-time Employee | 6-12 months | Psychology Masters Degree | 1,001-10,000 |
| S7 | Volunteer | 6-12 months | Psychology Masters Degree | <100 |
| S8 | Full-time Employee | 6-12 months | Psychology Masters Degree | 501-1000 |
| I1 | Volunteer | Less than 6 months | Psychology Masters Student | <100 |
| I2 | Volunteer | Less than 6 months | Psychology Masters Degree | <100 |
| I3 | Volunteer | Less than 6 months | Psychology Masters Degree | 501-1000 |
| I4 | Volunteer | Less than 6 months | Psychology Bachelors Student | 100-500 |
| I5 | Volunteer | Less than 6 months | Psychology Masters Degree | <100 |
| I6 | Volunteer | Less than 6 months | Psychology Masters Degree | >10,000 |
| I7 | Volunteer | Less than 6 months | Psychology Masters Degree | 100-500 |
| I8 | Volunteer | 6-12 months | Psychology Masters Student | 501-1000 |
| I9 | Volunteer | 6-12 months | Psychology Bachelors Student | <100 |
| I10 | Volunteer | 6-12 months | Psychology Bachelors Student | 100-500 |
| I11 | Full-time Employee | 1-3 years | Psychology Masters Degree | 1,001-10,000 |
| I12 | Volunteer | 1-3 years | Psychology Masters Degree | 100-500 |
| I13 | Volunteer | 1-3 years | Psychology Masters Degree | <100 |
| I14 | Volunteer | 1-3 years | Psychology Masters Degree | 100-500 |
| I15 | Volunteer | 1-3 years | Psychology Masters Degree | 100-500 |
| I16 | Volunteer | 1-3 years | Psychology Masters Degree | 100-500 |
| I17 | Full-time Employee | More than 3 years | Psychology Masters Degree | 501-1000 |
| I18 | Volunteer | More than 3 years | Psychology Masters Degree | <100 |
| I19 | Full-time Employee | More than 3 years | Psychology Masters Degree | 100-500 |
| I20 | Full-time Employee | More than 3 years | Psychology Masters Degree | 1,001-10,000 |

**Supplementary Table 1. Demographic and usage characteristics of study participants (n=28).** Participants include 8 counselors who were shadowed during counseling sessions (S1–S8) and 20 counselors who completed semi-structured interviews (I1–I20). Employment status, experience level, and educational background are self-reported. Copilot generation counts reflect the cumulative number of AI-generated suggestions requested by each counselor from their first use through March 16, 2026. Experience levels ranged from less than 6 months to more than 3 years, and educational backgrounds spanned from current bachelor's students to MPhil holders in psychology. The majority of participants were volunteers (n=20) with master's-level training in psychology.

# Appendix B: Codebook Summary

## Theme Overview

Our thematic analysis employed a comprehensive codebook with 5 major themes, 21 primary codes, and 75 subcategories to analyze counselor experiences with AI Copilot integration in crisis counseling.

## Theme 1: Acceptance, Adoption & Usage Patterns

*Understanding patterns of AI Copilot adoption and use among counselors*

- **1.1 Usage Status**: Current relationship with Copilot (Active User, Former User, Never User, Alternative AI User, Advanced Feature User)
- **1.2 Individual Differences in Adoption**: Factors influencing usage (Generational Differences, Educational Background, Personal Philosophy)
- **1.3 Contextual Triggers for Use**: Situations prompting Copilot activation (Feeling Stuck, More Empathy Needed, Phrasing Assistance, Specific Information/Resources Needed, Cognitive Overload, Time Pressure)
- **1.4 Reasons for Non/Limited Use**: Barriers to adoption (Skill Development Concerns, Different Workflow, Ethical Concerns, Copilot Sounds Robotic/AI)
- **1.5 Client Detection of AI & Impact**: Client awareness and responses (Detection Frequency, Detection Triggers, Reduction of Copilot Usage, Text Response, Voice Response)

## Theme 2: Copilot Functionality & Features

*Technical capabilities and limitations of the AI system*

- **2.1 Helpful Functions**: Valued capabilities (Paraphrasing, Empathy Generation, Phrasing/Structure, Rapport Building, Language Translation/Understanding)
- **2.2 Limitations & Technical Issues**: System shortcomings (Technical Problems, Context Blindness, Repetitive Responses, Solution-Focused Approach, Language Complexity Mismatch)
- **2.3 High Effectiveness Cases**: Situations where Copilot excels (Suicidality/NSSI, Anxiety, Grief/Loss)
- **2.4 Low Effectiveness Cases**: Situations where Copilot struggles (Suicidality/NSSI, Interpersonal Issues)
- **2.5 Requested Features & Improvements**: Desired enhancements (More Human-like Responses, Language Support, Style Adaptation, Better Context Understanding, Summary Generation)

## Theme 3: Changes to Work Practices

*How AI integration transforms counseling workflows*

- **3.1 Workflow Changes**: Process modifications (Efficiency Changes, Decision-Making Process, No Workflow Change)
- **3.2 Impact on Professional Skills**: Effects on counselor capabilities (Skill Enhancement, Protocol Reinforcement, Catching Missed Elements, Skill Atrophy Concerns, Transition Challenges)
- **3.3 Impact on Therapeutic Relationship**: Effects on counselor-client connection (Positive Impact, Negative Impact, No Impact)
- **3.4 Impact on Stress Management**: Effects on emotional labor (Positive Impact, Negative Impact, No Impact)

## Theme 4: Editing & Adaptation Practices

*How counselors modify and personalize AI suggestions*

- **4.1 Editing Frequency**: How often modifications occur (Always/Usually Edit, Sometimes Edit, Rarely Edit)
- **4.2 Types of Edits**: Nature of modifications (Question Removal/Modification, Tone Adjustment, Length Changes, Personal Style)
- **4.3 Editing Process**: Methods of modification (In-Tool Editing, Copy-Paste Method)

## Theme 5: Professional Identity & Values

*Impact on counselor identity and ethical considerations*

- **5.1 Core Professional Values**: Fundamental beliefs (Human Touch Essential)
- **5.2 Identity Threats & Concerns**: Challenges to professional self-concept (Replacement Anxiety, Dependency Fears)
- **5.3 Ethical Concerns**: Moral considerations (Authenticity, Confidentiality/Privacy, Environmental Impact)
- **5.4 Future of Counseling**: Visions of AI's role (AI as Assistant, Human Irreplaceability, AI Dominance Concerns)

## Appendix C: Inter-rater Agreement Improvements

| Metric | Before | After | Change |
|---|---|---|---|
| Mean Fleiss Kappa | 0.438 | 0.914 | 0.476 |
| Median Fleiss Kappa | 0.448 | 0.941 | 0.493 |
| SD Fleiss Kappa | 0.223 | 0.094 | -0.128 |
| Min Fleiss Kappa | -0.048 | 0.576 | 0.624 |
| Max Fleiss Kappa | 1.000 | 1.000 | 0.000 |
| Mean Percent Agreement | 68.6% | 94.6% | 26.0% |
| Median Percent Agreement | 69.0% | 96.6% | 27.6% |
| Codes with $\kappa < 0.4$ | 31 | 0 | -31 |
| Codes with Perfect Agreement | 3 | 21 | 18 |

**Supplementary Table 2. Per-code inter-rater reliability before and after the consensus process.** Fleiss' kappa and percent agreement were calculated for each thematic code across all raters. Before consensus, mean kappa indicated moderate agreement ($\kappa$=0.438), with 31 codes falling below the acceptable threshold ($\kappa$<0.4). After the two-stage consensus process, mean kappa rose to 0.914 (near-perfect agreement), all codes exceeded $\kappa$=0.576, and 21 codes achieved perfect agreement. Standard deviation of kappa decreased, reflecting more uniform reliability across codes.

| Metric | Before | After | Change |
|---|---|---|---|
| Mean Fleiss Kappa | 0.482 | 0.919 | 0.437 |
| Median Fleiss Kappa | 0.494 | 0.945 | 0.451 |
| Mean Jaccard Similarity | 0.483 | 0.916 | 0.433 |
| Median Jaccard Similarity | 0.483 | 0.939 | 0.456 |
| Mean Krippendorff Alpha | 0.482 | 0.919 | 0.437 |
| Problem Documents ($\kappa < 0.4$) | 5 | 0 | -5 |

**Supplementary Table 3.** Per-document inter-rater reliability before and after the consensus process. Agreement metrics were calculated for each of the 29 coded documents (20 interviews, 8 shadowing sessions, and the survey file) across all thematic codes. Before consensus, mean Fleiss' kappa (0.482), Jaccard similarity (0.483), and Krippendorff's alpha (0.482) all indicated moderate agreement, with 5 documents falling below $\kappa$=0.4. After consensus, all three metrics exceeded 0.91, and no documents remained below the acceptable reliability threshold.

# Appendix D: Extended Results Tables

| Survey Responses (n=56) | | | | | | |
|---|---|---|---|---|---|---|
| | **Non-psychology** | | **Psychology Undergraduate** | | **Psychology Post-Graduate** | |
| Educational Qualification | 8.93% | | 8.93% | | 82.14% | |
| | | | | | | |
| | **< 6 months** | **6 -12 months** | **1 - 3 years** | **> 3 years** | | |
| Overall Counselling Experience | 35.71% | 23.21% | 28.57% | 12.50% | | |
| | | | | | | |
| | **Extremely Unhelpful** | **Unhelpful** | **Somewhat Helpful** | **Helpful** | **Extremely Helpful** | **Mean** |
| 1. How would you rate the overall helpfulness of the AI Copilot tool? | 0.00% | 0.00% | 1.79% | 66.07% | 32.14% | 4.30 |
| | | | | | | |
| | **Extremely Unlikely** | **Unlikely** | **Neutral** | **Likely** | **Extremely Likely** | **Mean** |
| 2. If it is up to you, how likely are you to continue using the Copilot as it is right now? | 0.00% | 1.79% | 8.93% | 62.50% | 26.79% | 4.14 |
| | | | | | | |
| | **Strongly Disagree** | **Disagree** | **Neutral** | **Agree** | **Strongly Agree** | **Mean** |
| 4. The Co-Pilot allows me to ensure clients' physical safety in Crisis Counselling | 1.79% | 5.36% | 25.00% | 53.57% | 14.29% | 3.73 |
| 5. The Co-Pilot helps me to provide emotional stability for the clients during the moment of crisis | 1.79% | 3.57% | 21.43% | 53.57% | 19.64% | 3.86 |
| 6. The Co-Pilot helps me to improve my overall efficiency as a crisis counsellor | 0.00% | 1.79% | 17.86% | 51.79% | 28.57% | 4.07 |
| 7. The Co-Pilot helps me to structure and frame my sentences better during the counselling sessions | 0.00% | 1.79% | 14.29% | 37.50% | 46.43% | 4.29 |
| 8. The transition from manual crisis | 1.79% | 12.50% | 32.14% | 46.43% | 7.14% | 3.45 |
| counselling to using the Co-Pilot aligns with my counselling approach. | | | | | | |
| 9. Using the Co-Pilot has enhanced my counselling skills. | 0.00% | 7.14% | 23.21% | 53.57% | 16.07% | 3.79 |
| 10. Using the Co-Pilot has reduced my stress levels during crisis conversations. | 0.00% | 3.57% | 17.86% | 50.00% | 28.57% | 4.04 |
| 11. The Co-Pilot is easy to use during active conversations. | 0.00% | 7.14% | 12.50% | 64.29% | 16.07% | 3.89 |
| | | | | | | |
| | **No** | | | **Yes** | | |
| 12. Do you have any ethical concerns with using the Co-Pilot? | 85.71% | | | 14.29% | | |
| | | | | | | |
| | **Basic Prompt** | | **Advance Prompt** | | **Additional instruction** | |
| 14. Which specific features of the Co-Pilot do you find most helpful in improving your counselling effectiveness? | 55.36% | | 25.00% | | 19.64% | |

**Supplementary Table 4**. Summarized survey responses from counselors who used the AI Copilot during live crisis counseling conversations (n = 56). Means represent average scores on a 5-point Likert scale; mean values are not reported for the binary ethics item. Questions 3, 13, 15, and 16 of the survey were open-ended and their responses were analyzed as part of the thematic analysis.

| TAM2 Construct | Evidence from Our Data |
|---|---|
| **Core TAM Constructs** | |
| Perceived Usefulness | 98.2% (survey) rated the copilot as helpful or extremely helpful. Multiple dimensions of perceived value were demonstrated. |
| Perceived Ease of Use | 80.4% (survey) agreed or strongly agreed that the copilot is easy to use during active conversations. |
| Intention to Continue Use | 89.3% (survey) reported being likely or extremely likely to continue using the copilot. |
| **Cognitive Instrumental Processes** | |
| Job Relevance | The copilot was developed specifically to assist crisis counselors with their primary job tasks. Counselors have adopted it in a variety of situations; for example, 90% (interview) reported using it when feeling stuck, 85% (interview) when experiencing cognitive overload, and 85% (interview) for phrasing assistance. |
| Output Quality | Counselors praised the copilot's capabilities, including phrasing, empathy, catching missed details, and bringing new perspectives. However, at times, counselors encounter limitations like repetitiveness, context-blindness, solution-focused approaches, technical issues, and an AI-like voice. Counselors also edited messages, including removing/editing questions, adding/removing empathy statements, and adding specificity. |
| Result Demonstrability | 80.4% (survey) agreed the copilot improves efficiency, 78.6% (survey) reported stress reduction, and 70% (interview) described catching missed clinical elements, providing concrete examples of measurable benefits. |
| **Social Influence Processes** | |
| Subjective Norm | Leadership at the organization encourages use of the copilot through training and support, though no mandatory usage requirements or pressure were observed. |
| Voluntariness | Adoption is voluntary. |
| Image | No evidence of copilot use affecting peer status or professional standing among counselors. |
| **Novel Factors Beyond TAM2** | |
| Professional Development | 69.6% (survey) agreed the Copilot enhanced their counseling skills, while 5.4% (survey) and 55.0% (interview) expressed concerns about skill atrophy, with some counselors choosing non-adoption specifically to preserve skill development opportunities. |
| Authenticity | 10.7% (survey) reported concerns about authenticity, while 70% (interview) reduced Copilot usage after clients questioned whether they were speaking with AI, suggesting that authenticity concerns significantly influence usage patterns. |
| Ethical Values | 14.3% (survey) reported ethical concerns about using the copilot, with specific concerns including confidentiality (7.1% survey, 50% interview) and environmental impact (one counselor cited this as their primary reason for limited adoption). |

**Supplementary Table 5:** This table shows how data from interviews (n = 20), shadowing sessions (n = 8), and surveys (n = 56) map onto the constructs of the Technology Acceptance Model 2 (TAM2). It includes core TAM constructs (perceived usefulness, perceived ease of use, and intention to continue use), cognitive instrumental processes (job relevance, output quality, and result demonstrability), social influence processes (subjective norm, voluntariness, and image), as well as novel factors beyond TAM2 (professional development, authenticity, and ethical values), with supporting evidence drawn from counselor reports across all three data sources.

## Appendix E: Results of Thematic Analysis

| Theme Category | Interview (n=20) | Shadow (n=8) | Survey (n=56) |
|---|---|---|---|
| **1.1 Usage Status** | | | |
| Active User | 75.0% | 100.0% | 3.6% |
| Advanced Feature User | 15.0% | 25.0% | 0.0% |
| Alternative AI User | 80.0% | 0.0% | 0.0% |
| Former User | 25.0% | 12.5% | 1.8% |
| Never User | 25.0% | 12.5% | 0.0% |
| **1.2 Individual Differences in Adoption** | | | |
| Educational Background | 80.0% | 0.0% | 3.6% |
| Generational Differences | 40.0% | 0.0% | 0.0% |
| Personal Philosophy | 50.0% | 25.0% | 0.0% |
| **1.3 Contextual Triggers for Use** | | | |
| Cognitive Overload | 85.0% | 50.0% | 14.3% |
| Feeling Stuck | 90.0% | 75.0% | 19.6% |
| More Empathy Needed | 60.0% | 62.5% | 5.4% |
| Phrasing Assistance | 85.0% | 37.5% | 17.9% |
| Specific Information/Resources Needed | 25.0% | 37.5% | 0.0% |
| Time Pressure | 60.0% | 75.0% | 17.9% |

| Theme Category | Interview (n=20) | Shadow (n=8) | Survey (n=56) |
|---|---|---|---|
| **1.4 Reasons for Non/Limited Use** | | | |
| Copilot Sounds Robotic/AI | 75.0% | 37.5% | 21.4% |
| Different Workflow | 45.0% | 37.5% | 5.4% |
| Ethical Concerns | 20.0% | 12.5% | 1.8% |
| Skill Development Concerns | 60.0% | 25.0% | 3.6% |
| **1.5 Client Detection of AI & Impact** | | | |
| Detection Frequency | 95.0% | 37.5% | 3.6% |
| Detection Triggers | 95.0% | 50.0% | 16.1% |
| Reduction of Copilot Usage | 70.0% | 12.5% | 0.0% |
| Text Response | 70.0% | 37.5% | 0.0% |
| Voice Response | 40.0% | 25.0% | 0.0% |

**Supplementary Table 6.** Counselor acceptance and usage patterns of the AI Copilot across interviews (n = 20), shadowing sessions (n = 8), and survey responses (n = 56; unprompted mentions in open-ended answers). Themes capture variability in usage status, individual differences in adoption, contextual triggers for use, reasons for limited or non-use, and client detection of AI involvement.

| Theme Category | Interview (n=20) | Shadow (n=8) | Survey (n=56) |
|---|---|---|---|
| **2.1 Helpful Functions** | | | |
| Empathy Generation | 90.0% | 75.0% | 28.6% |
| Language Translation/Understanding | 40.0% | 12.5% | 3.6% |
| Paraphrasing | 60.0% | 37.5% | 25.0% |
| Phrasing/Structure | 85.0% | 62.5% | 16.1% |
| Rapport Building | 60.0% | 37.5% | 14.3% |
| **2.2 Limitations & Technical Issues** | | | |
| Context Blindness | 40.0% | 62.5% | 17.9% |
| Language Complexity Mismatch | 20.0% | 25.0% | 5.4% |
| Repetitive Responses | 55.0% | 62.5% | 23.2% |
| Solution-Focused Approach | 45.0% | 37.5% | 10.7% |
| Technical Problems | 20.0% | 25.0% | 8.9% |
| **2.3 High Effectiveness Cases** | | | |
| Anxiety | 5.0% | 25.0% | 8.9% |
| Grief/Loss | 0.0% | 0.0% | 3.6% |
| Suicidality or Non-suicidal self-injury (NSSI) | 60.0% | 50.0% | 19.6% |
| **2.4 Low Effectiveness Cases** | | | |
| Interpersonal Issues (Relationship/Family) | 0.0% | 0.0% | 5.4% |
| Suicidality or Non-suicidal self-injury (NSSI) | 20.0% | 50.0% | 14.3% |

| Theme Category | Interview (n=20) | Shadow (n=8) | Survey (n=56) |
|---|---|---|---|
| **2.5 Requested Features & Improvements** | | | |
| Better Context Understanding | 35.0% | 12.5% | 25.0% |
| Language Support | 10.0% | 0.0% | 1.8% |
| More Human-like Responses | 60.0% | 12.5% | 19.6% |
| Style Adaptation | 45.0% | 25.0% | 14.3% |
| Summary Generation | 50.0% | 25.0% | 10.7% |

**Supplementary Table 7.** Perceived functions, limitations, and improvement requests for the AI Copilot across interviews (n = 20), shadowing sessions (n = 8), and survey responses (n = 56; unprompted mentions in open-ended answers).

| Theme Category | Interview (n=20) | Shadow (n=8) | Survey (n=56) |
|---|---|---|---|
| **3.1 Workflow Changes** | | | |
| Decision-Making Process | 65.0% | 75.0% | 0.0% |
| Efficiency Changes | 85.0% | 75.0% | 16.1% |
| No Workflow Change | 55.0% | 0.0% | 0.0% |
| **3.2 Impact on Professional Skills** | | | |
| Catching Missed Elements | 70.0% | 0.0% | 0.0% |
| Protocol Reinforcement | 30.0% | 0.0% | 0.0% |
| Skill Atrophy Concerns | 55.0% | 12.5% | 5.4% |
| Skill Enhancement | 85.0% | 25.0% | 7.1% |
| Transition Challenges | 15.0% | 0.0% | 1.8% |
| **3.3 Impact on Therapeutic Relationship** | | | |
| Negative Impact | 50.0% | 12.5% | 5.4% |
| No Impact | 50.0% | 0.0% | 0.0% |
| Positive Impact | 55.0% | 0.0% | 0.0% |
| **3.4 Impact on Stress Management** | | | |
| Negative Impact | 5.0% | 0.0% | 0.0% |
| No Impact | 25.0% | 0.0% | 0.0% |
| Positive Impact | 70.0% | 0.0% | 5.4% |

**Supplementary Table 8:** Reported impacts of the AI Copilot on counselor workflow, skills, therapeutic relationships, and stress management, based on interviews (n = 20), shadowing (n = 8), and surveys (n = 56; unprompted mentions in open-ended answers).

| **Theme Category** | **Interview (n=20)** | **Shadow (n=8)** | **Survey (n=56)** |
|---|---|---|---|
| **4.1 Editing Frequency** | | | |
| Always/Usually Edit | 65.0% | 62.5% | 10.7% |
| Rarely Edit | 35.0% | 62.5% | 0.0% |
| Sometimes Edit | 50.0% | 87.5% | 1.8% |
| **4.2 Types of Edits** | | | |
| Length Changes | 60.0% | 37.5% | 8.9% |
| Personal Style | 85.0% | 75.0% | 5.4% |
| Question Removal/Modification | 65.0% | 62.5% | 8.9% |
| Tone Adjustment | 55.0% | 62.5% | 1.8% |
| **4.3 Editing Process** | | | |
| Copy-Paste Method | 60.0% | 50.0% | 0.0% |
| In-Tool Editing | 50.0% | 37.5% | 0.0% |

**Supplementary Table 9.** Counselor editing practices when using the AI Copilot, based on interviews (n = 20), shadowing (n = 8), and survey responses (n = 56; unprompted mentions in open-ended answers).

| Theme Category | Interview (n=20) | Shadow (n=8) | Survey (n=56) |
|---|---|---|---|
| **5.1 Core Professional Values** | | | |
| Human Touch Essential | 95.0% | 25.0% | 5.4% |
| **5.2 Identity Threats & Concerns** | | | |
| Dependency Fears | 65.0% | 12.5% | 7.1% |
| Replacement Anxiety | 30.0% | 12.5% | 5.4% |
| **5.3 Ethical Concerns** | | | |
| Authenticity | 90.0% | 12.5% | 10.7% |
| Confidentiality/Privacy | 50.0% | 12.5% | 7.1% |
| Environmental Impact | 0.0% | 12.5% | 1.8% |
| **5.4 Future of Counseling** | | | |
| AI Dominance Concerns | 20.0% | 12.5% | 1.8% |
| AI as Assistant | 80.0% | 12.5% | 10.7% |
| Human Irreplaceability | 80.0% | 0.0% | 0.0% |

**Supplementary Table 10.** Counselors' reflections on professional values, identity, and the future of counseling in relation to the AI Copilot, across interviews (n = 20), shadowing (n = 8), and surveys (n = 56; unprompted mentions in open-ended answers). Note: one counselor was shadowed twice, on an early version of the Copilot, and after implementation of the final version, so both shadowing sessions were consolidated into one for analysis purposes.

| | Percentages | | | | Means | | | |
|---|---|---|---|---|---|---|---|---|
| **Metric** | **< 6 months (n=20)** | **6-12 months (n=13)** | **1-3 years (n=16)** | **> 3 years (n=7)** | **< 6 months** | **6-12 months** | **1-3 years** | **> 3 years** |
| Overall Helpfulness | 100.0% | 100.0% | 93.8% | 100.0% | 4.25 | 4.31 | 4.31 | 4.43 |
| Likelihood to Continue | 95.0% | 92.3% | 75.0% | 100.0% | 4.25 | 4.15 | 3.94 | 4.29 |
| Physical Safety | 45.0% | 84.6% | 75.0% | 85.7% | 3.40 | 4.00 | 3.88 | 3.86 |
| Emotional Stability | 75.0% | 76.9% | 62.5% | 85.7% | 3.80 | 4.15 | 3.56 | 4.14 |
| Efficiency | 85.0% | 76.9% | 75.0% | 85.7% | 4.10 | 4.08 | 4.00 | 4.14 |
| Sentence Framing | 80.0% | 84.6% | 81.2% | 100.0% | 4.30 | 4.38 | 4.12 | 4.43 |
| Values Alignment | 65.0% | 46.2% | 43.8% | 57.1% | 3.70 | 3.38 | 3.19 | 3.43 |
| Skills Enhancement | 75.0% | 76.9% | 62.5% | 57.1% | 3.90 | 3.77 | 3.75 | 3.57 |
| Stress Reduction | 85.0% | 69.2% | 87.5% | 57.1% | 4.15 | 4.00 | 4.12 | 3.57 |
| Ease of Use | 75.0% | 84.6% | 75.0% | 100.0% | 3.85 | 3.92 | 3.69 | 4.43 |
| Ethical Concerns | 15.0% | 7.7% | 18.8% | 14.3% | 0.15 | 0.08 | 0.19 | 0.14 |

**Supplementary Table 11. Survey responses stratified by counseling experience (n=56).** Percentages reflect the proportion of counselors within each experience category who agreed or strongly agreed with each survey item; means reflect average scores on a 5-point Likert scale (1=Strongly Disagree, 5=Strongly Agree). Overall helpfulness was consistently high across all experience levels (≥93.8%). Notable differences emerged in several domains: newer counselors (<6 months) reported higher stress reduction (85.0%) and skills enhancement (75.0%) compared to the most experienced group (>3 years; 57.1% for both), while more experienced counselors reported stronger perceptions of physical safety (85.7% vs. 45.0%) and ease of use (100.0% vs. 75.0%). Likelihood to continue was lowest among counselors with 1–3 years of experience (75.0%). Ethical concerns were low across all groups (7.7%–18.8%). For the ethical concerns item, means represent the proportion responding "Yes" to a binary question.

# Appendix F

## Theoretical Implications: Extending TAM2 for AI-Augmented Professional Practice

While TAM2 captured core adoption drivers, three emergent factors reveal its limitations for AI integration in human service professions, suggesting necessary theoretical extensions.

**Professional Development**
TAM2 focuses on immediate job performance, missing how counselors evaluate AI's impact on skill development. Unlike traditional tools that augment abilities, AI can potentially substitute for clinical judgment. The 55.0% (interview) and 5.4% (survey) expressing skill atrophy concerns and some counselors' deliberate non-adoption to preserve learning opportunities indicate that technology acceptance in training-intensive professions requires incorporating long-term capability development as a distinct construct.

**Authenticity**
TAM2's "image" construct addresses peer status but misses authenticity concerns. Our finding that 70% of interviewees reduced usage after client questioning reveals a critical gap: when technology threatens the human connection essential to therapeutic relationships, perceived usefulness becomes secondary. Acceptance models for human services need an authenticity construct capturing impact on client trust and therapeutic alliance, not just professional reputation among peers.

**Ethical Values**
TAM2 assumes utility-based adoption, but at least one counselor rejected AI on principle. Acceptance models for mission-driven professions need to incorporate such values-based criteria.

## Appendix G: Prompt for copilot message edit analysis

You are an expert evaluator reviewing an edited message compared to a drafted message. Your task is to categorize the edits made by the counselor from the LLM message into all of the categories listed below that apply.

**Categories:**
**counselor_more_specific**
The edited response references a specific piece of information mentioned by the patient or makes a specific follow-up question whereas the draft message does not include details.

Example:
Draft message: I can only imagine how it must feel not to have that support system when you're going through this. Thank you for sharing this with me, Nila. I'm here to support you, please feel free to share more about your thoughts and feelings
Edited response: okay, I understand that it makes it difficult to share with your friends as they didn't want you to get into another relationship because of what happened before. Thank you for sharing this with me, Nila. I'm here to support you

**counselor_less_specific**
The edited response makes the message broader and more general for the patient.
Example:

**counselor_modifies_question**
The edited response has a different or modified question from the drafted message.
Example:
Draft message: It's understandable to struggle with feelings of guilt and confusion, especially when comparing yourself to others. I'm here to support you; what would you like to focus on as a first step?
Edited response: It's understandable to struggle with feelings of guilt and confusion with everything that has happened, but you do not have to stay with these difficult feelings, would you like to tell me if you have been doing anything to manage these feelings?

**counselor_removes_question**
The edited reponse does not have a question whereas the draft message includes a question to explore the situation further.
Example:
Draft message: I can only imagine how overwhelming it must be to feel blamed all the time and not have a safe space to express yourself. I am here to support you and listen to what's on your mind. What would you like to focus on or talk about next?
Edited response: I can only imagine how overwhelming it must be to feel blamed all the time and not have a safe space to express yourself. I am here to support you and listen

**counselor_less_empathy**
Edited response contained fewer empathy statements (support, validation). The draft. uses more sentences of comfort.
Example:
Draft message: It sounds like you're managing a lot on your own, and I understand it can be distressing without the support you need. Have you had a chance to discuss this situation with someone who might help you communicate your desire to leave this job?
Edited response: Have you had a chance to discuss this situation with someone who might help you communicate your desire to leave this job?

**counselor_more_empathy**
Edited response contained additional empathetic statements (support, validation). The draft used less sentences of comfort.
Example
Draft message: I can imagine how painful it is not to have said goodbye, especially when you were so close to your grandfather. It's natural to feel this way, and it may take time to process. With everything going on, is there someone you feel comfortable talking to, like a friend or family member?
Edited response: I can imagine how painful it is not to have said goodbye, especially when you were so close to your grandfather. It must be so hurtful. It's natural to feel this way, and it may take time to process. With everything going on, is there someone you feel comfortable talking to, like a friend or family member?"

**Other**
All edits that do not fall under the categories listed above will go under the Other category.

**Instructions for Output:**
- Carefully analyze the original message from the LLM and the edited message from counselor
- Categorize each edit to all of the categories it applies to (every edit should be under at least one category).
- Provide a brief explanation for the categorization.
- Output the evaluation in JSON format with a justification.

**Output Format Example:**

```json
 {{
   "category": ["counselor_modifies_question", "counselor_more_specific"],
     "justification": "The counselor generally guided the conversation using more specific questions that included personal details without having to reconfirm the patient’s feelings."
}}
```

Now, evaluate the original message from the LLM and the edited message from the counselor and provide your assessment.

{note}

# Appendix H: Recruitment and Sampling Procedures

| Data Source | Recruitment Process | Number Invited | Number Responded | Leadership involved in recruitment? |
|---|---|---|---|---|
| **Interviews** | Counselors were asked to be interviewed and shadowed in a non-random manner to cover a wide range of demographic categories: Employment Status (Volunteer/Full-time), Experience Level (from < 6 months to > 3 years), Educational Background (from Student to Masters Degree), and Copilot Usage Level (# Copilot generations, from <100 to > 10,000). | 20 | 20 | Yes |
| **Shadowing** | | 8 | 8 | Yes |
| **Survey** | All counselors who have used the copilot for a month were sent the survey in April and May 2025. | 69 | 56 | Yes |